# Parallel Decoders of Polar Codes

Bin Li, Hui Shen, and David Tse, *Fellow, IEEE*

*Abstract*—In this letter, we propose parallel SC (Successive Cancellation) decoder and parallel SC-List decoder for polar codes. The parallel decoder is composed of $M = 2^m (m \geq 1)$ component decoders working in parallel and each component decoder decodes a Polar code of a block size of $1/M$ of the original Polar code. Therefore the parallel decoder has $M$ times faster decoding speed. Our simulation results show that the parallel decoder has almost the same error-rate performance as the conventional non-parallel decoder.

## I. INTRODUCTION

Polar codes are a major breakthrough in coding theory [1]. They can achieve Shannon capacity with a simple encoder and a simple successive cancellation decoder, both with low complexity of the order of $O(N \log N)$, where $N$ is the code block size. When the code block size is long enough, the simple SC decoder can approaches Shannon capacity. But for short and moderate lengths, the error rate performance of polar codes with the SC decoding is not as good as LDPC or turbo codes. A new SC-list decoding algorithm was proposed for polar codes recently [2], which performs better than the simple SC decoder and performs almost the same as the optimal ML (maximum likelihood) decoding at high SNR. In order to improve the low minimum distance of polar codes, the concatenation of polar codes with simple CRC was proposed [2], and it was shown that a simple concatenation scheme of polar code (2048, 1024) with a 16-bit CRC using the SC-List decoding can outperform Turbo and LDPC codes [3][4].

Although Polar codes provide good error-rate performance, the SC and SC-List decoder work in a serial fashion. They decode information bits one-by-one. It is very hard to achieve a high decoding speed or low latency due to this serial decoding. Some work [5][6] has been on reducing the decoding latency of SC decoder of Polar codes by optimizing the hardware design of the SC decoder, and the improvement in the decoding speed is 2 times faster. In this letter, we propose both parallel SC and SC-List decoder to overcome this drawback. This parallel decoder consists of many component decoders which work in parallel, therefore we can achieve much higher decoding speed or much lower latency. Our improvement in the decoding speed is $M = 2^m (m \geq 1)$ times faster if $M$ component decoders are used. Simulations show that the parallel decoders provide the same error-rate performance as the conventional decoders.

In section II, we propose three parallel decoders with two component decoders, with four component decoders and with eight component decoders. We also provide simulation results. Finally we draw some conclusions in section III.

## II. PARALLEL SC AND SC-LIST DECODERS

### A. Polar Codes

Let $F = \begin{bmatrix} 1 & 0 \\ 1 & 1 \end{bmatrix}$, $F^{\otimes n}$ is a $N \times N$ matrix, where $N = 2^n$, $\otimes n$ denotes $n$th Kronecker power, and $F^{\otimes n} = F \otimes F^{\otimes (n-1)}$. Let the $n$-bit binary representation of integer $i$ be $b_{n-1}, b_{n-2}, ..., b_0$. The $n$-bit representation $b_0, b_1, ..., b_n$ is a bit-reversal order of $i$. The generator matrix of polar code is defined as $G_N = B_N F^{\otimes n}$, where $B_N$ is a bit-reversal permutation matrix. The polar code is generated by

$$x_1^N = u_1^N G_N = u_1^N B_N F^{\otimes n} \quad (1)$$

where $x_1^N = (x_1, x_2, ..., x_N)$ is the encoded bit sequence, and $u_1^N = (u_1, u_2, ..., u_N)$ is the encoding bit sequence. The bit indexes of $u_1^N$ are divided into two subsets: the one containing the information bits and the other containing the frozen bits. For simplicity, the frozen bits are set "0".

### B. Parallel SC Decoder with Two Component Decoders

Due to the special structure of $F^{\otimes n}$, Polar code can be expressed as

$$\begin{aligned} x_1^N &= u_1^N B_N \times \begin{bmatrix} F^{\otimes(n-1)} & 0 \\ F^{\otimes(n-1)} & F^{\otimes(n-1)} \end{bmatrix} \\ &= \begin{bmatrix} v_1^{N/2} B_{N/2} & v_{N/2+1}^N B_{N/2} \end{bmatrix} \times \begin{bmatrix} F^{\otimes(n-1)} & 0 \\ F^{\otimes(n-1)} & F^{\otimes(n-1)} \end{bmatrix} \end{aligned} \quad (2)$$

where $u_1^N B_N = \begin{bmatrix} v_1^{N/2} B_{N/2} & v_{N/2+1}^N B_{N/2} \end{bmatrix}$

Furthermore, we have

$$x_1^N = \begin{bmatrix} a_1^{N/2} B_{N/2} F^{\otimes(n-1)} & b_1^{N/2} B_{N/2} F^{\otimes(n-1)} \end{bmatrix} \quad (3)$$

where $a_1^{N/2} = v_1^{N/2} \oplus v_{N/2+1}^N$ and $b_1^{N/2} = v_{N/2+1}^N$.

From (3), we can see that one Polar code of block size $N$ can be decomposed into two sub Polar codes each with a block size of $N/2$, but the encoding bits $a_1^{N/2}$ and $b_1^{N/2}$ are correlated.

Bin Li and Hui Shen are with the Communications Technology Research Lab., Huawei Technologies, Shenzhen, P. R. China (e-mail:{binli.binli, hshen}@huawei.com).

David Tse is with the Dept. of Electrical Engineering and Computer Science, University of California at Berkeley, CA 94720-1170, USA (e-mail: dtse@eecs.berkeley.edu).

After passing $x_1^N$ through a channel, we have the received signal $y_1^N$. We propose a parallel SC decoder to decode the received signal as follows. The parallel SC decoder consists of two component SC decoders: one uses $y_1^{N/2}$ as input to decode $a_1^{N/2}$, and the other uses $y_{N/2+1}^N$ as input to decode $b_1^{N/2}$. Two SC decoders calculate log likelihood ratios: $L_{N/2}^{(i)}(y_1^{N/2}, \hat{a}_1^{i-1})$ and $L_{N/2}^{(i)}(y_{N/2+1}^N, \hat{b}_1^{i-1})$ independently as follows

$$L_{N/2}^{(i)}(y_1^{N/2}, \hat{a}_1^{i-1}) = \log \frac{W_{N/2}^{(i)}(y_1^{N/2}, \hat{a}_1^{i-1}|a_i=0)}{W_{N/2}^{(i)}(y_1^{N/2}, \hat{a}_1^{i-1}|a_i=1)} \quad (4)$$

$$L_{N/2}^{(i)}(y_{N/2+1}^N, \hat{b}_1^{i-1}) = \log \frac{W_{N/2}^{(i)}(y_{N/2+1}^N, \hat{b}_1^{i-1}|b_i=0)}{W_{N/2}^{(i)}(y_{N/2+1}^N, \hat{b}_1^{i-1}|b_i=1)} \quad (5)$$

The decision on $a_i$ and $b_i$ is made either independently or jointly. If both $v_i$ and $v_{N/2+i}$ are information bits, then $a_i = v_i \oplus v_{N/2+i}$ and $b_i = v_{N/2+i}$ are independent from each other and decisions are made independently as

$$\hat{a}_i = \begin{cases} 0, & \text{if } L_{N/2}^{(i)}(y_1^{N/2}, \hat{a}_1^{i-1}) \geq 0 \\ 1, & \text{otherwise} \end{cases} \quad (6)$$

$$\hat{b}_i = \begin{cases} 0, & \text{if } L_{N/2}^{(i)}(y_{N/2+1}^N, \hat{b}_1^{i-1}) \geq 0 \\ 1, & \text{otherwise} \end{cases} \quad (7)$$

If both $v_i$ is a frozen bit and $v_{N/2+i}$ is a information bit, then $a_i = b_i = v_{N/2+i}$, an equal gain combing of the log likelihood ratios is used to decode:

$$\hat{a}_i = \begin{cases} 0, & \text{if } L_{N/2}^{(i)}(y_1^{N/2}, \hat{a}_1^{i-1}) + L_{N/2}^{(i)}(y_{N/2+1}^N, \hat{b}_1^{i-1}) \geq 0 \\ 1, & \text{otherwise} \end{cases} \quad (8)$$

Fig. 1 shows the parallel decoder structure. Since the two component SC decoders operate decoding in parallel, the decoding speed of this parallel decoder is two times faster than the conventional SC decoder.

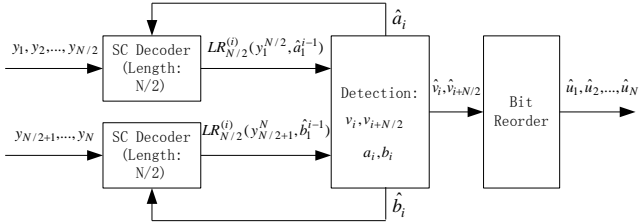

Fig. 1 The Proposed Parallel SC Decoder with Two Component Decoders.

## C. Parallel SC Decoder with Four Component Decoders

The Polar code can be further expressed as

$$x_1^N = u_1^N B_N \times \begin{bmatrix} F^{\otimes(n-2)} & 0 & 0 & 0 \\ F^{\otimes(n-2)} & F^{\otimes(n-2)} & 0 & 0 \\ F^{\otimes(n-2)} & 0 & F^{\otimes(n-2)} & 0 \\ F^{\otimes(n-2)} & F^{\otimes(n-2)} & F^{\otimes(n-2)} & F^{\otimes(n-2)} \end{bmatrix} \quad (9)$$

Let $u_1^N B_N = \begin{bmatrix} v_1^{N/4} B_{N/4} & v_{N/4+1}^{N/2} B_{N/4} & v_{N/2+1}^{3N/4} B_{N/4} & v_{3N/4+1}^N B_{N/4} \end{bmatrix}$, we have

$$\begin{cases} x_1^{N/4} = a_1^{N/4} B_{N/4} F^{\otimes(n-2)} \\ x_{N/4+1}^{N/2} = b_1^{N/4} B_{N/4} F^{\otimes(n-2)} \\ x_{N/2+1}^{3N/4} = c_1^{N/4} B_{N/4} F^{\otimes(n-2)} \\ x_{3N/4+1}^N = d_1^{N/4} B_{N/4} F^{\otimes(n-2)} \end{cases} \quad (10)$$

where

$$\begin{cases} a_i = v_i \oplus v_{i+N/4} \oplus v_{i+N/2} \oplus v_{i+3N/4} \\ b_i = v_{i+N/4} \oplus v_{i+3N/4} \\ c_i = v_{i+N/2} \oplus v_{i+3N/4} \\ d_i = v_{i+3N/4} \end{cases} \quad 1 \leq i \leq N/4 \quad (11)$$

From (10), we can see that one Polar code of a block size $N$ can be decomposed into four sub Polar codes each with a block size of $N/4$, but the encoding bits $a_1^{N/4}$, $b_1^{N/4}$, $c_1^{N/4}$ and $d_1^{N/4}$ are correlated.

A proposed parallel SC decoder consists of four component decoders as shown in Fig. 2. The four component decoders use $y_1^{N/4}$, $y_{N/4+1}^{N/2}$, $y_{N/2+1}^{3N/4}$ and $y_{3N/4+1}^N$ as inputs to decode $a_1^{N/4}$, $b_1^{N/4}$, $c_1^{N/4}$ and $d_1^{N/4}$, respectively, and they calculate log likelihood ratios: $L_{a_i} = L_{N/4}^{(i)}(y_1^{N/4}, \hat{a}_1^{i-1})$, $L_{b_i} = L_{N/4}^{(i)}(y_{N/4+1}^{N/2}, \hat{b}_1^{i-1})$, $L_{c_i} = L_{N/4}^{(i)}(y_{N/2+1}^{3N/4}, \hat{c}_1^{i-1})$ and $L_{d_i} = L_{N/4}^{(i)}(y_{3N/4+1}^N, \hat{d}_1^{i-1})$ independently. Since $(\hat{a}_i, \hat{b}_i, \hat{c}_i, \hat{d}_i)$ are correlated, we will firstly decode independent bits $(v_i, v_{i+N/4}, v_{i+N/2}, v_{i+3N/4})$, which are decoded by maximizing the equally combined log likelihood ratios:

$$\{\hat{v}_i, \hat{v}_{i+N/4}, \hat{v}_{i+N/2}, \hat{v}_{i+3N/4}\} = \arg\{ \max_{v_i, v_{i+N/4}, v_{i+N/2}, v_{i+3N/4}} [(1-2a_i)L_{a_i} + (1-2b_i)L_{b_i} + (1-2c_i)L_{c_i} + (1-2d_i)L_{d_i}]\} \quad (12)$$

where $(a_i, b_i, c_i, d_i)$ is a function of $(v_i, v_{i+N/4}, v_{i+N/2}, v_{i+3N/4})$ defined in (11).

Since $v_1^N$ is an interleaved version of $u_1^N$, $(v_i, v_{i+N/4}, v_{i+N/2}, v_{i+3N/4})$ may contain frozen bits which are set "0", the maximization of (12) is an exhaustive search of all combinations of the non-frozen bits. After $(v_i, v_{i+N/4}, v_{i+N/2}, v_{i+3N/4})$ is detected, we obtain $(\hat{a}_i, \hat{b}_i, \hat{c}_i, \hat{d}_i)$ from (11). If $(v_i, v_{i+N/4}, v_{i+N/2}, v_{i+3N/4})$ are all information bits, then $(a_i, b_i, c_i, d_i)$ are independent from each other and the above detection can be simplified as detecting $(a_i, b_i, c_i, d_i)$ independently according to their log-likelihood ratios.

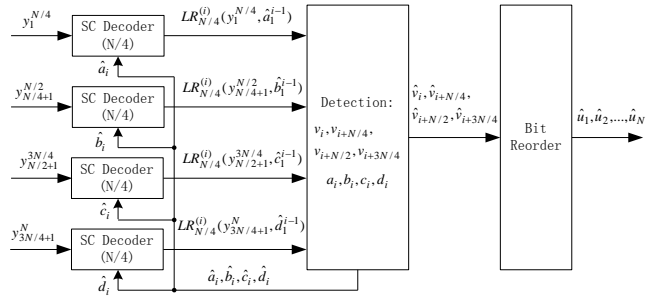

Fig. 2 The Proposed Parallel SC Decoder with Four Component Decoders.

## D. Parallel SC Decoder with Eight Component Decoders

Straightforwardly, let

$$u_1^N B_N = [v_1^{N/8} B_{N/8} \; v_{N/8+1}^{2N/8} B_{N/8} \; \cdots \; v_{7N/8+1}^{N} B_{N/8}] \quad (13)$$

We have

$$\begin{cases} x_1^{N/8} = a_1^{N/8} B_{N/8} F^{\otimes(n-3)} \\ x_{N/8+1}^{2N/8} = b_1^{N/8} B_{N/8} F^{\otimes(n-3)} \\ \vdots \\ x_{7N/8+1}^{N} = h_1^{N/8} B_{N/8} F^{\otimes(n-3)} \end{cases} \quad (14)$$

where

$$\begin{cases} a_1^{N/8} = v_1^{N/8} \oplus v_{N/8+1}^{2N/8} \oplus \cdots \oplus v_{6N/8+1}^{7N/8} \oplus v_{7N/8+1}^{N} \\ b_1^{N/8} = v_{N/8+1}^{2N/8} \oplus v_{3N/8+1}^{4N/8} \oplus v_{5N/8+1}^{6N/8} \oplus v_{7N/8+1}^{N} \\ c_1^{N/8} = v_{2N/8+1}^{3N/8} \oplus v_{3N/8+1}^{4N/8} \oplus v_{6N/8+1}^{7N/8} \oplus v_{7N/8+1}^{N} \\ d_1^{N/8} = v_{3N/8+1}^{4N/8} \oplus v_{7N/8+1}^{N} \\ e_1^{N/8} = v_{4N/8+1}^{5N/8} \oplus v_{5N/8+1}^{6N/8} \oplus v_{6N/8+1}^{7N/8} \oplus v_{7N/8+1}^{N} \\ f_1^{N/8} = v_{5N/8+1}^{6N/8} \oplus v_{7N/8+1}^{N} \\ g_1^{N/8} = v_{6N/8+1}^{7N/8} \oplus v_{7N/8+1}^{N} \\ h_1^{N/8} = v_{7N/8+1}^{N} \end{cases} \quad (15)$$

From (14), we can obtain a parallel SC decoder with eight component SC decoders as shown in Fig. 3.

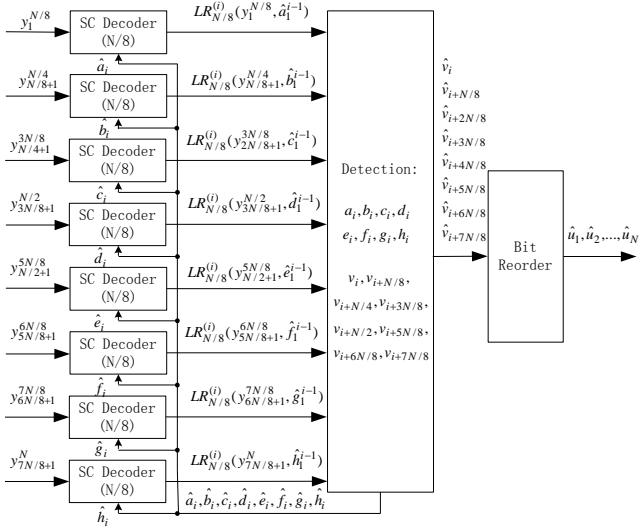

Fig. 3 The Proposed SC decoder with Eight Component Decoders.

## E. Parallel SC-LIST Decoder with Two Component Decoders

We propose a parallel SC-LIST decoder to decode the received signal as shown in Fig. 4. The parallel SC decoder consists of two component SC-LIST decoder: decoder A uses $y_1^{N/2}$ as input to decode $a_1^{N/2}$, and decoder B uses $y_{N/2+1}^{N}$ as input to decode $b_1^{N/2}$. Each path $P_m = \{A_m, B_m\}$ $(1 \le m \le L)$ consists of two sub-paths: $A_m = \{\hat{a}_{m,1}, \hat{a}_{m,2}, \cdots, \hat{a}_{m,k-1}\}$ for the decoder A and $B_m = \{\hat{b}_{m,1}, \hat{b}_{m,2}, \cdots, \hat{b}_{m,k-1}\}$ for the decoder B. At decoding time $k$, the decoder A generates $2L$ new paths: $\{A_m, \hat{a}_{m,k}=0\}$ and $\{A_m, \hat{a}_{m,k}=1\}$, $(1 \le m \le L)$, and the decoder B generates $2L$ new paths: $\{B_m, \hat{b}_{m,k}=0\}$ and $\{B_m, \hat{b}_{m,k}=1\}$ $(1 \le m \le L)$. We generate $L/2L/4L$ combined paths according to $(v_k, v_{k+N/2})$ as follows: 1) If both $v_k$ and $v_{k+N/2}$ are frozen bits, then $\hat{a}_k = \hat{b}_k = 0$ and $L$ combined paths are $P_m = \{A_m, B_m, \hat{a}_{m,k}=0, \hat{b}_{m,k}=0\}$, $(1 \le m \le L)$; These $L$ paths are split into two groups: $\{A_m, \hat{a}_{m,k}=0\}$ $(1 \le m \le L)$ for the decoder A and $\{B_m, \hat{b}_{m,k}=0\}$ $(1 \le m \le L)$ for the decoder B. 2) If $v_k$ is a frozen bit and $v_{k+N/2}$ is a information bit, then $\hat{a}_k = \hat{b}_k = v_{k+N/2}$ and we generate $2L$ combined paths: $\{A_m, B_m, \hat{a}_{m,k}=0, \hat{b}_{m,k}=0\}$ and $\{A_m, B_m, \hat{a}_{m,k}=1, \hat{b}_{m,k}=1\}$ $(1 \le m \le L)$; The path metrics of the combined path $\{A_m, B_m, \hat{a}_{m,k}, \hat{b}_{m,k}\}$ is the sum of path metrics of path $\{A_m, \hat{a}_{m,k}\}$ and path $\{B_m, \hat{b}_{m,k}\}$. 3) If both $v_k$ and $v_{k+N/2}$ are information bits, $4L$ combined paths are generated as $\{A_m, B_m, \hat{a}_{m,k}=v_k \oplus v_{k+N/2}, \hat{b}_{m,k}=v_{k+N/2}\}$, where $v_k, v_{k+N/2} \in \{0,1\}$; If the number of combined paths is larger than a predefined threshold $L_{\max}$, $L_{\max}$ best combined paths are kept. We choose $L_{\max} = L$ in Fig. 4. Each combined path is split into two sub-paths with one for the decoder A and one for the decoder B.

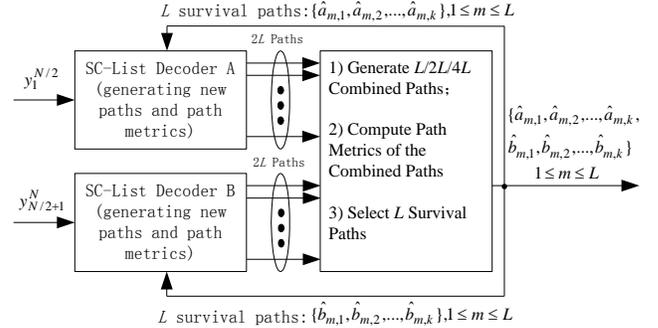

Fig. 4 The Proposed SC-LIST decoder with Two Component Decoders.

## F. Parallel SC-LIST Decoder with Four Component Decoders

We propose a parallel SC-LIST decoder to decode the received signal as shown in Fig. 5. The parallel SC decoder consists of four component SC-LIST decoders A/B/C/D: They use $y_1^{N/4}$, $y_{N/4+1}^{N/2}$, $y_{N/2+1}^{3N/4}$ and $y_{3N/4+1}^{N}$ as inputs to decode $a_1^{N/4}$, $b_1^{N/4}$, $c_1^{N/4}$ and $d_1^{N/4}$, respectively. Each path $P_m = \{A_m, B_m, C_m, D_m\}$ $(1 \le m \le L)$ consists of four sub-paths: $A_m = \{\hat{a}_{m,1}, \hat{a}_{m,2}, \cdots, \hat{a}_{m,k-1}\}$, $B_m = \{\hat{b}_{m,1}, \hat{b}_{m,2}, \cdots, \hat{b}_{m,k-1}\}$, $C_m = \{\hat{c}_{m,1}, \hat{c}_{m,2}, \cdots, \hat{c}_{m,k-1}\}$ and $D_m = \{\hat{d}_{m,1}, \hat{d}_{m,2}, \cdots, \hat{d}_{m,k-1}\}$ for the decoder A, B, C and D, respectively. At decoding time $k$, each decoder produces $2L$ new paths. We generate $L/2L/4L/8L/16L$ combined paths according to all combinations by taking different values of information bits in $(v_k, v_{k+N/4}, v_{k+N/2}, v_{k+3N/4})$, and for each combination, $\{\hat{a}_{m,k}, \hat{b}_{m,k}, \hat{c}_{m,k}, \hat{d}_{m,k}\}$ is calculated as follows: $\hat{a}_k = \hat{v}_k \oplus \hat{v}_{k+N/4} \oplus \hat{v}_{k+N/2} \oplus \hat{v}_{k+3N/4}$, $\hat{b}_k = \hat{v}_{k+N/4} \oplus \hat{v}_{k+3N/4}$, $\hat{c}_k = \hat{v}_{k+N/2} \oplus \hat{v}_{k+3N/4}$ and $\hat{d}_k = \hat{v}_{k+3N/4}$. Let $w$ be the number of information bits in $(v_k, v_{k+N/4}, v_{k+N/2}, v_{k+3N/4})$, the number of the generated combined paths is $2^w L$. Since $w \in \{0,1,2,3,4\}$, $2^w \in \{1,2,4,8,16\}$. If $2^w L > L_{\max}$, the $L_{\max}$ best paths are kept, and these $L_{\max}$ paths are split into $4L_{\max}$ sub-paths for the decoder A, B, C and D. We choose $L_{\max} = L$ in Fig. 5.

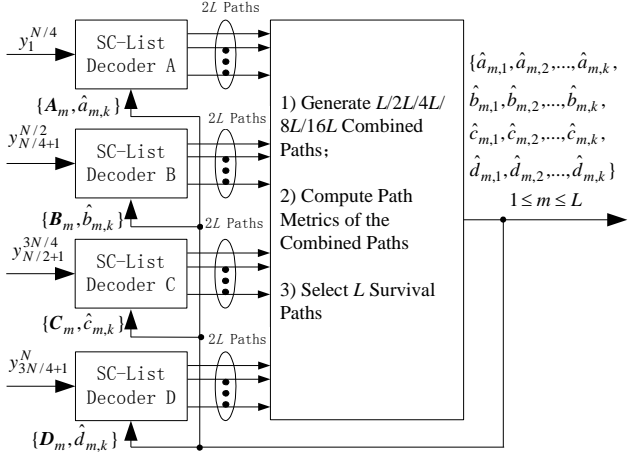

Fig. 5 The Proposed SC-LIST decoder with Four Component Decoders.

### G. Performance Simulations

Fig. 6 shows bit error rate (BER) and frame error rate (FER) of Polar code (2048,1024) with the conventional SC decoding and our proposed parallel SC decoding. Fig. 7 shows BER and FER of Polar code (2048,1008) with 16-bit CRC using the conventional SC-LIST and our proposed parallel SC-LIST decoding, where the list size is $L_{\max} = 32$ and the adaptive list algorithm[4] is used. It is shown that the parallel decoders perform almost the same as the conventional non-parallel decoder for both the SC and the SC-LIST decoding.

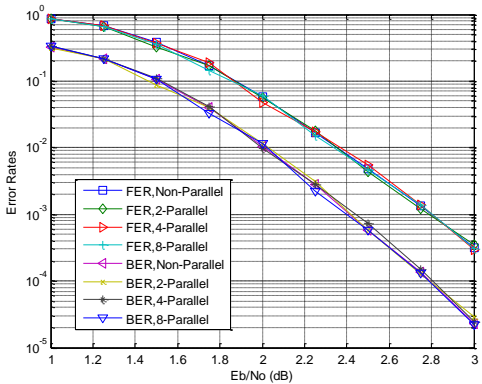

Fig. 6 The error-rate performance of polar code (2048, 1024) with the conventional SC and our proposed SC parallel decoding.

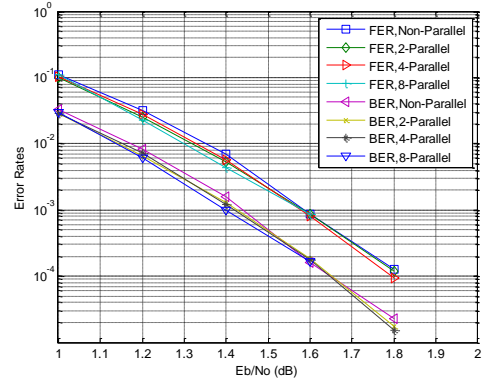

Fig. 7 The error rates of polar code (2048, 1008) with 16-bit CRC using the conventional SC-List and our proposed parallel SC-List decoding.

### III. CONCLUSION

In this letter, we propose parallel SC and SC-LIST decoder which are composed of $M = 2^m (m \geq 1)$ component decoders working in parallel. The proposed parallel decoder can provide $M$ times faster decoding speed than the conventional decoder. Since each component decoder decodes a Polar code with a block size $N/M$, the decoding complexity of each component decoder is in order of $O(N/M \log(N/M))$ for the SC decoder and in order of $O(LN/M \log(N/M))$ for the SC-LIST decoder. The overall complexity of the whole parallel decoder is in order of $O(N \log(N/M))$ for the SC decoder and in order of $O(LN \log(N/M))$ for the SC-LIST decoder, which is less than the conventional SC and the SC-LIST decoder, respectively. Our simulation results show that the parallel decoders perform almost the same as the conventional non-parallel decoders.

### REFERENCES

[1] E. Arıkan, "Channel polarization: A method for constructing capacity achieving codes for symmetric binary-input memoryless channels," IEEE Trans. Inform. Theory, vol. 55, pp. 3051–3073, July 2009.
[2] I. Tal and A. Vardy, "List Decoding of Polar Codes," available as online as arXiv: 1206.0050v1.
[3] E. Arıkan, "Polar Coding: Status and Prospects", Plenary Talk of IEEE International Symposium on Inform. Theory, Saint Petersburg, Russia, 2011.
[4] B. Li, H. Shen, and D. Tse, "An Adaptive Successive Cancellation List Decoder for Polar Codes with Cyclic Redundancy Check," IEEE Comm. Letters, vol. 16, pp. 2044–2047, Dec. 2012.
[5] C. Leroux, I. Tal, A. Vardy, and W. J. Gross, "Hardware Architectures for Successive Cancellation Decoding of Polar Codes", IEEE ICASSP, 2011.
[6] C. Zhang, B. Yuan, and K. K. Parhi, " Reduced-latency SC polar decoder architectures," ICC 2012，pp. 3471 – 3475.